\documentclass[preprint,review,10pt]{elsarticle}
\biboptions{authoryear}

\usepackage{array}
\usepackage{paralist}
\usepackage{enumitem}
\setitemize{noitemsep,topsep=0pt,parsep=0pt,partopsep=0pt}

\usepackage{graphicx}
\usepackage{xcolor}
\usepackage{url}
\usepackage{amsmath, amsthm, amssymb}
\usepackage{booktabs}

\usepackage{makecell}

\newcommand{\ie}{\textit{i.e.}, }
\newcommand{\eg}{\textit{e.g.}, }

\def\footnoterule{\hrule \kern 3pt}

\biboptions{round,sort,comma}

\begin{document}

\begin{frontmatter}

\title{New centrality and causality metrics assessing air traffic network interactions}

\author[a]{Piero Mazzarisi}
\author[a]{Silvia Zaoli\fnref{label1}}
\author[a]{Fabrizio Lillo}
\author[b]{Luis Delgado}
\author[b]{G\'erald Gurtner}

\address[a]{Dipartimento di Matematica, University of Bologna, Bologna, Italy}
\address[b]{School of Architecture and Cities, University of Westminster, London, United Kingdom}

\fntext[label1]{To whom correspondence should be addressed. E-mail: silvia.zaoli@unibo.it}

\begin{abstract}

In ATM systems, the massive number of interacting entities makes it difficult to identify critical elements and paths of disturbance propagation, as well as to predict the system-wide effects that innovations might have. To this end, suitable metrics are required to assess the role of the interconnections between the elements and complex network science provides several network metrics to evaluate the network functioning. Here we focus on centrality and causality metrics measuring, respectively, the importance of a node and the propagation of disturbances along links. By investigating a dataset of US flights, we show that existing centrality and causality metrics are not suited to characterise the effect of delays in the system. We then propose generalisations of such metrics that we prove suited to ATM applications. Specifically, the new centrality is able to account for the temporal and multi-layer structure of ATM network, while the new causality metric focuses on the propagation of extreme events along the system.

\end{abstract}

\end{frontmatter}

\newpage

\section{Introduction}
\label{sec:intro}

The introduction of changes in the ATM system is often difficult due to the tight interdependencies that exists across the different systems, subsystems, and institutional frameworks. The full implications of changes on parts of the system are difficult to predict at system level. 

At a time of increased traffic, the ATM system can improve its performance by being better tuned for flexibility. For example, understanding the coupling between flights helps to understand the margins embedded into the flight schedules designed by airlines and can lead to better knowledge of the coupling between stakeholders and processes.

Current monitoring of ATM performances are based on \emph{classical} indicators (KPIs) which are estimated considering different stakeholders. However, the interdependencies among the system elements are not adequately represented. Capturing these interdependencies is critical in order to understand the current system performances and how changes affect the relationship between the elements in the system. This can be mitigated with the use of network metrics \citep{zaninLillo} such as \emph{centrality} and \emph{causality}, which quantify the network connectivity and highlight the delay propagation patterns. Moreover, it is relevant to consider how system's elements are connected and how their criticality to propagate delay and cost might be different from the perspective of different stakeholders (and in particular for flights and passengers) \citep{POEM, Montlaur2017}. \\
The purpose of this paper is twofold. First, we show that existing network metrics are not suitable to identify the effect of delays and missed connections in the functioning of the ATM system at a global and local level. Second, we propose new metrics which are tested on real traffic data and are designed to capture loss of airport centralities and to identify channels of disturbance propagation. By using suitable simulation models, these network metrics can be used to assess the effect of the introduction of new mechanisms (e.g. modifiying the flight arrival coordination) on the ATM system. Within the Domino project we have developed a detailed Agent Based Model simulating the whole European airspace. One of the objectives of the project has been exactly to use the new metrics presented here to test the effect of three different innovations in the ATM. The model and the results of the application of the new metrics to its output will be presented in a separate article. \\ 
The paper is organized as follows. Section \ref{sec:background} presents the background on metrics for the estimation of ATM performances. The network metrics of centrality and causality are described in Section~\ref{sec:network_metrics}, along with their application to a US traffic dataset. This helps to highlight the potential and limitation of these metrics. 
Finally, in Section~\ref{sec:conclusions} we draw some conclusions.

\section{Background -- classical metrics}
\label{sec:background}

When analysing the performance of the ATM system, a set of metrics are usually used in the ATM community. These can be grouped by different areas and stakeholders. Some of them capture the interaction of elements in the system but in an implicit manner, as the network view of the system is not explicitly represented. It is common to consider average values for the metrics even if it has been shown that their distribution are critical to understand the system performance. This is particularly relevant in the case of delay and cost of delay due to the non-linearity between them \citep{cost_delay}.

SESAR identifies 6 different key performance areas (KPAs) with different key performance indicator (KPIs) that need to be monitored in order to assess the impact of introducing different solutions. Table~\ref{t:sesar_indicators} summarises these.

\begin{table}[t]
\begin{center}
\caption{SESAR performance KPAs/KPIs (adapted from ~\cite{master_plan2015})}
\label{t:sesar_indicators}
\begin{tabular}{p{0.25\textwidth}|p{0.75\textwidth}}
\toprule
Key performance area & Key performance indicator\\
\hline
Cost efficiency: ANS productivity & \begin{minipage}[t]{\linewidth}
   \begin{itemize}[noitemsep,topsep=0pt,parsep=-5pt,partopsep=0pt]
    \item Gate-to-gate direct ANS cost per flight
    
    \begin{itemize}[noitemsep,topsep=0pt,parsep=-5pt,partopsep=0pt]
        \item Determined unit cost for en-route ANS
        \item Determined unit cost for terminal ANS
    \end{itemize}

\end{itemize} \end{minipage}\\
Operational efficiency & \begin{minipage}[t]{\linewidth}
    \begin{itemize}[noitemsep,topsep=0pt,parsep=-5pt,partopsep=0pt]
    \item Fuel burn per flight (tonne/flight)
    \item Flight time per flight (min/flight)
\end{itemize} \end{minipage}\\
Capacity & \begin{minipage}[t]{\linewidth}
    \begin{itemize}[noitemsep,topsep=0pt,parsep=-5pt,partopsep=0pt]
    \item Departure delay (min/dep)
    \item En-route air traffic flow management delay
    \item Primary and reactionary delays all causes
    \item Additional flights at congested airports (million)
    \item Network throughput additional flights (million)
\end{itemize} \end{minipage}\\
Environment & \begin{minipage}[t]{\linewidth}
    \begin{itemize}[noitemsep,topsep=0pt,parsep=-5pt,partopsep=0pt]
    \item $CO_{2}$ emissions (tonne/flight)
   
    \begin{itemize}[noitemsep,topsep=0pt,parsep=-5pt,partopsep=0pt]
        \item Horizontal flight efficiency (actual trajectory)
        \item Vertical efficiency
        \item Taxi-out phase
    \end{itemize}
\end{itemize} \end{minipage}\\

Safety & \begin{minipage}[t]{\linewidth}
    \begin{itemize}[noitemsep,topsep=0pt,parsep=-5pt,partopsep=0pt]
    \item Accidents with ATM contribution
\end{itemize} \end{minipage}\\
Security & \begin{minipage}[t]{\linewidth}
    \begin{itemize}[noitemsep,topsep=0pt,parsep=-5pt,partopsep=0pt]
    \item ATM related security incidents resulting in traffic disruptions
\end{itemize}  \end{minipage}

\end{tabular}
\end{center}
\end{table}

These indicators allow stakeholders to monitor the performance of the system at a very high level and to define political goals. When considering the full impact of introducing new solutions in the system, one should consider the different stakeholders and the trade-offs that emerge between lower level indicators. In particular metrics should be defined for:
\begin{itemize}
    \item ANSPs
    \item Airports
    \item Airspace users
    \item Passengers
    \item Environment
\end{itemize}

Table~\ref{t:basic_stakeholder} summarises, per stakeholder, different metrics that have been considered in previous research \citep{cass_II,vista_final}.

\begin{table}[t]
\begin{center}
\caption{Basics metrics per stakeholder}
\label{t:basic_stakeholder}
\begin{tabular}{p{0.2\textwidth}|p{0.6\textwidth}}
\toprule

Stakeholder & Metrics\\
\hline

ANSP & \begin{minipage}[t]{\linewidth}
    \begin{itemize}[noitemsep,topsep=0pt,parsep=-5pt,partopsep=0pt]
    \item En-route airspace charges revenues
\end{itemize} \end{minipage} \\
Airport & \begin{minipage}[t]{\linewidth}
    \begin{itemize}[noitemsep,topsep=0pt,parsep=-5pt,partopsep=0pt]
    \item departing queue delay
    \item arrival queue delay
    \item number of operations
    \begin{itemize}[noitemsep,topsep=0pt,parsep=-5pt,partopsep=0pt]
        \item departures
        \item arrivals
    \end{itemize} 
\end{itemize} \end{minipage}\\
Airspace users & \begin{minipage}[t]{\linewidth}
    \begin{itemize}[noitemsep,topsep=0pt,parsep=-5pt,partopsep=0pt]
    \item flight departure delay
    \item flight arrival delay
    \item fuel
    \item delay per flight segment
    \item reactionary delay
    \item ATFM delay
    \item gate-to-gate time
    \item cost of delay
    
    \begin{itemize}[noitemsep,topsep=0pt,parsep=-5pt,partopsep=0pt]
        \item non-passenger related
        \item passenger related (hard and soft)
    \end{itemize} 
    
    \item cost
    \begin{itemize}
        \item en-route charges
        \item fuel cost
    \end{itemize}
    
\end{itemize} \end{minipage}\\
Passengers & \begin{minipage}[t]{\linewidth}
    \begin{itemize}[noitemsep,topsep=0pt,parsep=-5pt,partopsep=0pt]
    \item departure delay
    \item arrival delay
    \item missed connections
    \item connecting time
    \item gate-to-gate time
\end{itemize} \end{minipage} \\
Environment & \begin{minipage}[t]{\linewidth}
    \begin{itemize}[noitemsep,topsep=0pt,parsep=-5pt,partopsep=0pt]
    \item fuel kg
    \item $CO_{2}$ tonnes
\end{itemize}
\end{minipage}

\end{tabular}
\end{center}
\end{table}

It has been pointed out several times how similar metrics (\eg delay) could be experienced very differently by different stakeholders and in particular the differences between flight-centric and passenger-centric metrics \citep{cass_II}. For example, reductions in flight arrival delay with passenger arrival delay map close to a 1:1.3 ratio \citep{vista_final}. That is, on average, one minute of flight delay corresponds to 1.3 minutes of delay per passenger. This is due to the fact that the delay experienced by passengers is higher due to missed connections \citep{vista_final}.

This is one of the main reasons why when analysing the system performance not only flight-centric but also passenger-centric metrics should be considered. This duality might be relevant also when considering the interaction between elements in the system at network level.

These metrics are useful for performance monitoring and to understand the impact of different ATM solutions on the different stakeholders and their trade-offs. However, they do not address the complexity of the network nor provide information on how the different elements are related in the system. For this reason, specific ad hoc network metrics should be considered.

\section{Network metrics}
\label{sec:network_metrics}

Centrality and causality metrics are network metrics which can be applied to the ATM system, as this can be seen as a network whose nodes are the airports. In this section, we define and apply them to a dataset of 2015 US flights with the aim of  highlighting their capabilities and limitations.

Centrality is a measure of the importance of a node in a network. While several different definitions of centrality exist, all centrality metrics are based on some concept of connectivity of a node in terms of links, paths or walks joining it to the other nodes of the network. In the ATM network, we can consider the flights as links between the airports. Then, when airports are ranked according to an appropriate centrality measure, the airports with the highest ranks are the ones providing to the passengers the highest potential of moving through the network. The loss of centrality of an airport, between the scheduled and the realised network, signals a diminished potential of moving through the network passing through that node, which means, from the passenger's point of view, a diminished performance of the network. This loss of centrality should reflect both the missing links due to cancellations and the disrupted paths due to delays. Provided a centrality metric satisfying these requirements, comparing the loss of centrality between the realised and the scheduled flight network among case studies implementing different mechanisms would allow to assess the impact of innovations on the network performance. In particular, an innovation minimising the centrality losses between the scheduled and realised network represents an improvement from the passengers' point of view. In section \ref{sec:stateart_cent} we review some of the most commonly used centrality metrics and in section \ref{sec:data_cent} we show their limitations in describing the loss of connectivity of the network due to delays. Finally, in section \ref{sec:new_metrics_cent} we present a recently proposed centrality metrics suited for the air traffic network, Trip centrality \citep{Zaoli2019}, and show that it serves our purposes.

In the ATM system, delays and congestion states propagate through the system due to the entangled interactions between the flights and the environment, \eg the network manager, the airports or the arrival coordinators. As innovations aim to reduce the propagation of delays, the complex network toolbox should include a metric able to detect the extent to which the congested state of an airport causes congestion in other nodes of the network. In Time Series Analysis, a (directional) causal relation between two systems is detected when the information on the state of one system helps in predicting the future state of the other. The presence of a causal relation is assessed by means of statistical tests whose most famous example is the Granger causality metrics  \citep{granger1969investigating}. Indeed, it has been recently applied to airport networks \citep{Cook2015,zanin2017network}.
Here, a data driven approach is adopted to identify the channels through which the delay propagates and establish a network of causal relations, where a link between two airports is present if delay propagates from one to the other. Causality is tested between the states of congestion of airports in the network, measured as the average flight delay for that airport. 
The topology of the resulting causal network may change depending on the mechanism implemented in the system. This relates the presence of innovations at the micro level to its impact on delay dynamics and propagation at some macro level of aggregation, such as airports, airlines or passengers. For example, a smaller number of causal links and less causal feedbacks can be seen as an improvement of the system, as they signal a diminished coupling of the systems' elements.  In section \ref{sec:stateart_caus} we review Granger causality metrics and its recent application to ATM systems. Then, in section \ref{sec:data_caus} we show some limitations in describing non-linear aspects of delay propagation and possible spurious causal relations as a consequence of the autocorrelation structure of the delay states. Finally, in section \ref{sec:new_metrics_caus} we suggest the improvements that could be introduced to the existing metrics.

\subsection{Dataset}
\label{sec:dataset}
To show the limitations of existing metrics (centrality and causality), we apply them to the network of flights operated in 2015 by 14 major US airlines. The dataset was obtained from the U.S. Department of Transportation's (DOT) Bureau of Transportation Statistics. For each flight, the dataset reports the date, the airline operating it, the departure and arrival airport, the scheduled departure and arrival times and the realised ones, the aircraft tail number, whether it was cancelled or diverted. All schedules were converted from local time to Eastern Standard Time (EST). For the centrality analysis, performed on one day, the day was considered to start at 4AM EST. This choice reflects the fact that, as shown already in \citep{Fleurquin2013}, very few flights depart between 0AM an 4AM local time, therefore 4AM EST is a time of minimum activity across all the country. Causality analysis was instead performed on hourly time series ranging from one to three months.

\subsection{Centrality metrics}
\label{sec:centrality_metrics}
\subsubsection{State of the art}
\label{sec:stateart_cent}
Commonly used centrality metrics apply to single-layer static networks. Let us therefore start by considering the network of flights and airports aggregated across layers, \ie across airlines, and across time frames, \ie where all flights are present at the same time regardless of their schedule. Let $A$ be the weighted adjacency matrix of the network, such that $A_{ij}=k$ if there are $k$ flights going from $i$ to $j$. Here, we consider three among the most common and well known centrality metrics: degree centrality, Katz centrality, and Page Rank. Since the network of flights and airports is directed, a distinction should be made, in each case, between incoming and outgoing centrality.\\  
The incoming (outgoing) degree centrality of a node $i$ is given by the number of incoming (outgoing) edges (each flight is considered as an edge), 
\begin{equation}
    d^{IN}_i=\sum_j A_{ji},
\end{equation}
\begin{equation}
   d^{OUT}_i=\sum_j A_{ij}, 
\end{equation} 
where the index $j$ runs on all the nodes. This centrality metric measures with how many flights node $i$ can be reached (respectively, how many flights depart from node $i$). However, an important feature of the flight network are connections, which make use of two or more flights. A commonly used metric which considers a node's centrality to depend on the walks of any length arriving to (or departing from) that node is Katz centrality \citep{Newman2010}. The incoming Katz centrality of node $i$ is 
\begin{equation}
    k^{IN}_i= \sum_j (\mathbb{I}-\alpha A)^{-1}_{ji}=\sum_j \sum_{n=0}^\infty \alpha^n (A^n)_{ji},
\end{equation} 
where $\mathbb{I}$ is the identity matrix. Thus each walk of length $n$ from any node $j$ of the network to $i$ contributes $\alpha^n$ to the centrality of $i$. Since $\alpha<1$, longer walks contribute less and its value determines the contribution of long walks to centrality. The weight coefficient $\alpha$ must be smaller than the inverse of the largest eigenvalue of $A$ for the expression to converge \citep{Newman2010}. Correspondingly, the outgoing Katz centrality of node $i$ is 
\begin{equation}
    k^{OUT}_i= \sum_j (\mathbb{I}-\alpha A)^{-1}_{ij}=\sum_j \sum_{n=0}^\infty \alpha^n (A^n)_{ij}.
\end{equation}
 Page Rank is a generalisation of Katz centrality, developed by Google, that introduces an additional weight to the paths, depending on the in- (or out-) degree of the nodes they cross. Specifically, 
 \begin{equation}
     p^{IN}_i=\sum_j (\mathbb{I}-\alpha D^{-1}A)^{-1}_{ji},
 \end{equation}
 where $D_{ij}=\delta_{ij} d^{OUT}_j$, so that a link from $j$ to $k$ is weighted by the inverse of the out-degree of $j$, $1/d^{OUT}_j$.\\

\subsubsection{Application of the existing metrics to the US flights dataset}
\label{sec:data_cent}
\begin{figure}[t]
\begin{center}
\includegraphics{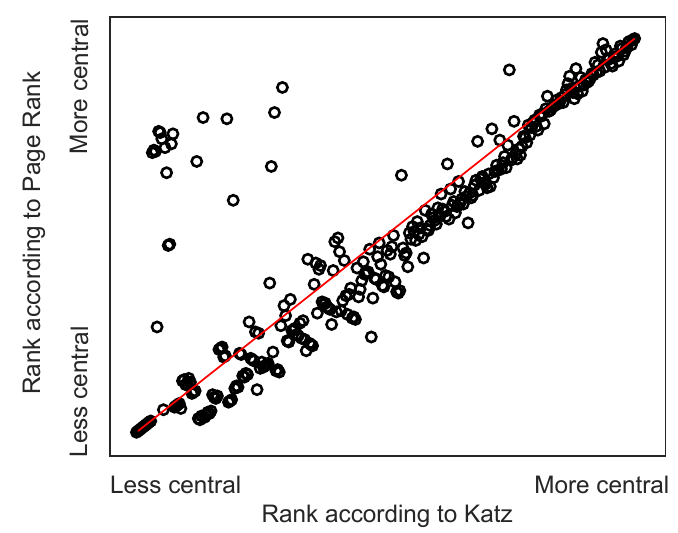}
\caption{Comparison of airport ranks according to incoming Katz centrality and incoming Page Rank centrality for the scheduled network on day 1. The red line is the 1:1 line. Points above the red line represent airports having gained importance with Page Rank. }
\label{fig:katzVSpr}
\end{center}
\end{figure}

To apply centrality metrics, we selected two days of the dataset differing in the amount of delay realised on the network. We considered four global parameters characterising delay: the fraction of delayed flights, the total delay, the average delay, and the average delay of delayed flights. On the first selected day, April 3rd 2015, all these parameters are below or close to the average (computed on all days), while on the second considered day, April 9th 2015, all parameters are above average. Additionally, April 3rd had 87 cancelled flights, while April 9th had 246.  In the following, we refer to these two days respectively as ``day 1" and ``day 2". For each day, we computed the airports' ranking according to each of the three centrality metrics reviewed in section \ref{sec:centrality_metrics}, incoming and outgoing, for the scheduled and the realised network. The obtained ranking are compared using the Kendall rank correlation coefficient $\tau$, which measures the similarity of two ranked sequences of data. The coefficient takes values in [-1,1], with the value 1 corresponding to two identical sequences and the value -1 to two sequences that are one the inverse of the other. \\
For Katz centrality, we chose $\alpha=0.003$, assuring convergence of the metric for both chosen days. Note that this small value of $\alpha$ strongly penalises long walks, therefore we do not expect the ranking to differ much from the degree ranking. For Page Rank centrality, instead, larger $\alpha$s still allow convergence, therefore we chose $\alpha=\exp(-1/2)$, so that walks of length $n\le 2$ are given a non negligible weight. \\ 
The rankings according to incoming and outgoing centralities result are very similar according to all three metrics, displaying, for day 1, respectively $\tau=$ 0.97, 0.97 and 0.93 on the scheduled network and $\tau=$ 0.97, 0.97 and 0.93 for the realised one. Also the rankings according to the centrality computed on the scheduled network and on the realised one are quite similar for both days. For day 1, the rankings display correlations, respectively for the three metrics, $\tau=$0.996, 0.995 and 0.995 in the incoming case and $\tau=$0.996, 0.991 and 0.991 in the outgoing case. For day 2, we have $\tau=$ 0.990, 0.985 and 0.995 in the incoming case and $\tau=$ 0.980, 0.976 and 0.992 in the outgoing case. The slightly smaller rank correlations coefficients for day 2 are due to the larger number of cancelled flights with respect to day 1. However, none of the considered centrality measures is able to reflect the fact that, on day 2, the much larger and abundant delays certainly caused more disruption of the network connectivity. In fact, if cancelled flights were excluded from the analysis (\ie they are not counted in the scheduled network either), these static metrics would not see any centrality loss at all due to the delays, therefore the rankings according to the scheduled and the realised networks would be identical. \\
While the rankings according to degree and Katz centrality are similar (for the scheduled network, incoming case, $\tau=$ 0.90 for day 1 and $\tau=$ 0.88 for day 2), Page Rank introduces stronger ranking differences with respect to Katz (for the scheduled network, incoming case, $\tau=$ 0.77 for day 1 and $\tau=$ 0.68 for day 2)\footnote{This difference is not due to the different values of $\alpha$ in the two cases.}. Figure \ref{fig:katzVSpr} shows a comparison of the two rankings, highlighting that most of the difference is due to a group of airports having a low ranking according to Katz centrality and getting a strong ranking boost with Page Rank (in the upper left part of the figure). These are mostly small airports in Alaska having direct flights to the airport of Anchorage. As Anchorage has itself a strong rank increase due to having several directed flights from airports with low out-degree, all the airports connected to it by a direct flight  also increase their ranking. This outcome, with a set of peripheral airports climbing the ranking, questions the suitability of Page Rank centrality to characterise node importance in ATM networks. In general, these differences between different centrality metrics highlight the fact that each metric describes a different aspect of the network structure, and care should be taken in their comparison. For example, degree considers only direct links, therefore it is appropriate if we are interested in assessing the potentiality of an airport to provide direct connections to other airports of the network, but it is not able to evaluate the role of flight connections. Katz centrality and Page Rank, instead, take into account also walks of any length on the network. While walks on the aggregated, static network considered here do not correspond to real itineraries that can be followed, accounting for longer walks means attributing centrality to an airport if it is connected to other central airports. Therefore, these two metrics are more appropriate when we want to assess the the potentiality of an airport to provide connections to other airports of the network with walks of any length. As a consequence of the different way of weighting walks in the two metrics, Katz centrality favours airports linked to large airports (with many link), as they will have many walks departing or arriving, while Page Rank rather tends to  favour airports with more links to smaller sized airport.

\subsubsection{A new centrality metric for the ATM system}
\label{sec:new_metrics_cent}
\begin{figure}[t]
\begin{center}
\includegraphics{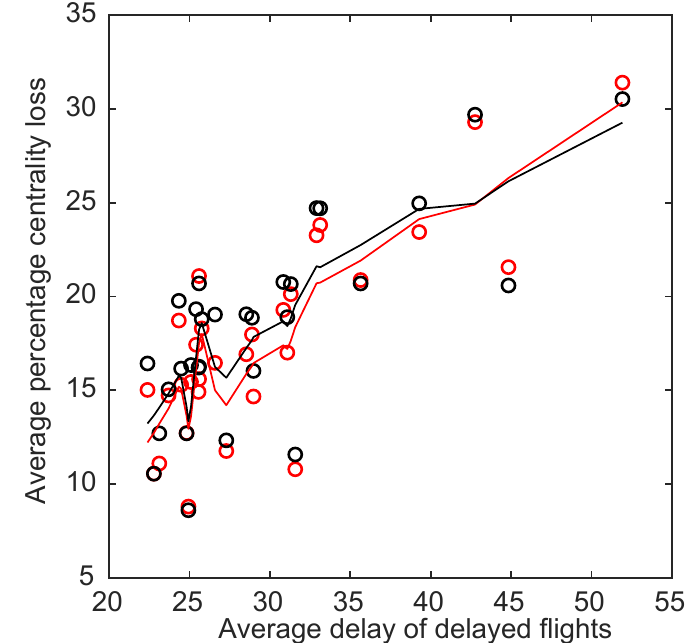}
\caption{Percentage of Trip Centrality loss, averaged over all airports, in each day of the dataset, according to incoming Trip Centrality (red) and outgoing Trip Centrality (black) plotted against the average departure delay of delayed flights in minutes. Trip centrality is computed with $\alpha=0.2$ and $\varepsilon=0$. Each point corresponds to one day of the dataset. The percentage centrality loss of an airport is computed as $\Delta c_{\%}= 100 \times (c_{sched}-c_{act})/c_{sched}$, where $c_{sched}$ and $c_{act}$ are the airport's centralities on the scheduled and realised network. Lines are obtained by a locally weighted smoothing (LOWESS) of the dots of the correspondent colour.}
\label{fig:TCvsDelay}
\end{center}
\end{figure}

To evaluate the effect of changes in the ATM system on the network performance, a centrality metric should be able to tell apart a situation where delays disrupt connections to one where they do not. Specifically, an airport's centrality should reflect its participation to walks that can actually be travelled, \ie respecting the schedule, so that disrupted connections imply a centrality drop. We showed in section \ref{sec:data_cent} that this is not the case for existing centrality metrics. In fact, all three metrics presented here do not account for the temporal structure of the network. Katz and Page Rank centrality, in particular, count walks on the network which are not time ordered and therefore have no relation with the trajectories that passengers could travel. As a consequence, these metrics cannot reflect the effect of delays on the network's connectivity. Additionally, the weight assigned to each walk does not consider which airline each flight composing the walk belongs to, therefore a walk using only flights of one airline has the same weight of a walk of the same length using several airlines. However, a more realistic assumption would be that the latter contributes less to centrality, as it is travelled with a smaller probability. Accounting for this requires considering the multiplex structure of the network.\\ 
Generalisations of the existing metrics should therefore be devised to overcome these limitations. A version of Katz centrality for temporal network has been proposed in \citep{Grindrod2011}. It considers adjacency matrices $A^{[t]}$ containing only the links present in a time frame around time $t$ and counts walks which are ordered in time. However, it does not account for the fact that a link's duration coincides with the time it takes to travel though that link, affecting the feasibility of a walk. A solution to account for link's schedule by introducing secondary nodes was introduced in \citep{Zanin2009}. By joining these two ideas, a new centrality metric, named ``Trip Centrality'' was proposed by three of us in \citep{Zaoli2019}. To compute Trip centrality, a secondary node is introduced for each link (\ie flight) in the network, and such link is substituted by two `stubs', one from the origin node to the secondary one, present only in the time frame during which the original link was appearing (time of departure) and one from the secondary node to the destination one, present only in the time frame during which the original link was disappearing (time of arrival). With the introduction of secondary nodes, the time of residence in a secondary node coincides with the time it takes to travel through the original link. If $\{A^{[t_i]}\}_{i=1,\dots T}$ are time-discretized adjacency matrices for the network with secondary nodes and stubs as described above, the time ordered products $A^{[t_1]}\dots A^{[t_n]}$ with $t_1<t_2<\dots <t_n$ count only the time-ordered walks using feasible link connections. Then, the $i,j$-th element of the matrix
\begin{equation}
\label{eq:Q}
    Q=[(\mathbb{I}+\tilde{\alpha} A^{[1]})(\mathbb{I}+\tilde{\alpha}  A^{[2]})\dots(\mathbb{I}+\tilde{\alpha}  A^{[T]})-\mathbb{I}],
\end{equation}
contains the contribution to centrality of all walks from $i$ to $j$ ($\tilde{\alpha}=\sqrt{\alpha}$ is the weight of a one-stub walk). The vectors of temporally generalised outgoing and incoming Katz centrality are then obtained summing $Q$, respectively. Further details can be found in \citep{Zaoli2019}.\\
Furthermore, to differentiate between within-airline and across-airlines walks, the multiplex nature of the network should be considered. Centrality measures for multiplexes are reviewed in \citep{Boccaletti2014}, however they either consist in computing the centrality of an airport separately on each layer and then aggregating the single-layer centralities to obtain a global centrality (\eg by summing or averaging the single-layer centralities) or in computing the centrality on an aggregated network, which adjacency matrix is the sum of the adjacency matrices of all layers. The first approach only counts within-airline walks, neglecting inter-layer ones. The second one, which corresponds to what we have presented in section \ref{sec:stateart_cent}, counts instead both intra- and inter-layer walks without distinction in weights. In Trip Centrality, a parameter $\epsilon\in [0,1]$ weights each change of layer, so that walks using links on several layers are included in the centrality computation but contribute less than an intra-layer walk of the same length. In the limit in which $\epsilon=0$, only intra-layer walks are counted, while in the limit $\epsilon=1$ no distinction is made between intra- and inter-layer walks. This weighting is obtained as follows (see \cite{Zaoli2019} for a more detailed explanation). For each primary node, one copy for each layer is considered, such that an adjacency matrix has size $(N N_L+N_l)\times(N N_L+N_l)$, where $N$ is the number of primary links (airports), $N_L$ is the number of layers (airlines) and $N_l$ is the number of secondary nodes (flights). Then,  we introduce the matrix $K$, of the same size of $A$, as the matrix with elements $K_{ii}=1$ and $K_{ij}=\varepsilon$ if $i$ and $j$ are two copies of the same node on different layers (zero otherwise). Now, the products of the form $A^{[t_1]}KA^{[t_2]}K\dots K A^{[t_n]}$ count time-ordered walks by introducing a factor $\varepsilon$ every time there is a change of layer. In conclusion, the outgoing Trip Centrality  on the temporal multiplex is written as
\begin{equation}
\label{eq:Trip}
  \vec{t}^{out}=[(\mathbb{I}+\tilde{\alpha} A^{[1]}K)(\mathbb{I}+\tilde{\alpha}  A^{[2]}K)\dots(\mathbb{I}+\tilde{\alpha}  A^{[T]}K)-\mathbb{I}]K^{-1}\vec{1}_{(N N_L+N_l)},
\end{equation}
where $\vec{1}_{(N N_L+N_l)}$ is a column vector of ones (used to perform the sum over columns).
The incoming centrality is generalised similarly. With this formula, a centrality for each copy of a node on each different layer is obtained. An aggregated centrality value fo a node, counting walks outgoing from (or incoming to) that node on any layer, is obtained by summing the centralities of all its copies. \\  
The outgoing Trip centrality of an airport counts all the walks, \ie "potential" passenger itineraries, having that airport as the origin, while the incoming Trip centrality counts those having that airport as a destination. Potential itineraries are all the sequences of any number of flights that can be potentially taken one after the other, given their schedule. An itinerary of $n$ legs is weighted $\alpha^n$, where $\alpha<1$ , so that itineraries made of more legs are counted less. Note that, due to how the metric is computed, no upper or lower limit for the connecting time is considered, so that two flights can be taken in sequence as long as the second one departs later than the arrival of the first.\\
Cancellations and delays make some of the walks that existed in the scheduled network not feasible anymore in the actual one. The resulting damage to the network connectivity can be quantified by the loss of centrality between the scheduled and the actual network. Centrality in the actual network is computed by using the actual network structure, which accounts for the delays and cancellations, and by excluding from the counting the new itineraries that become possible due to delays. \\
The application of Trip Centrality to the US dataset proves that this metric is able to capture the network effect of delays, differently from the static centrality metrics. In fact, Figure \ref{fig:TCvsDelay} plots the percentage centrality loss, averaged over all airports, for each day against the average delay of delayed flights on that day and shows an overall increasing patter. This means that centrality losses tend to be larger when delays are larger. In \citep{Zaoli2019} it was shown that the average centrality loss is also increasing with two other delay-related indicators: the average fraction of delayed flights in an airport and the average delay of all flights on that day. It was also shown that when these indicators increase the rankings in the scheduled and realised network tend to be more different. This remains true also when the cancelled flights are excluded from the analysis (see supplementary figure S3 of \citep{Zaoli2019}), proving that the effect of missed connection is recognised. \\
Differently from Katz centrality, in Trip centrality the parameter $\alpha$ weighting the use of one link can be chosen without constraints, because all counted walks  are made of a finite number of jumps (at most one per time-frame). For values of $\alpha$ large enough, say $\alpha>0.05$, the ranking according to Trip centrality differs significantly from the ones obtained by Katz or degree centrality (see \cite{Zaoli2019}), as such values of the parameter give importance to walks longer than one, on whose counting Trip Centrality differs. 

\subsection{Causality metrics} 
\label{sec:causality_metrics}
\subsubsection{State of the art}
\label{sec:stateart_caus}
A method to test whether there is a causal relation between two time series was first proposed by Granger \citep{granger1969investigating} and is based on the idea that, if the knowledge of past observations of one time series allows us to estimate future observations of the other time series better than without considering them, then there exists a directional causal relation. Here, we review the application of the Granger causality metrics to the ATM network system. We quantify an airport's congestion by a stochastic variable $X$ whose realisation $x_{t}$ at time $t$ is given by, for example, the average delay of flights taking off from that airport in the time interval centred in $[t,t+\Delta t]$. Flight delay is defined as the difference between realised and  scheduled departing times. We considered $\Delta t=$1 hour and when no departing flights are present in the interval we set $x_{t}=0$.

\paragraph{Granger causality in mean \citep{granger1969investigating}}
$Y\equiv\{y_t\}_{t=1,...,T}$ is said to Granger-cause $X\equiv\{x_t\}_{t=1,...,T}$ if we reject the null hypothesis that the past values of $Y$ do not provide statistically significant information about future values of $X$ by assuming VAR(p) as the predictive model \citep{johnston1972econometric}. Let us consider $X$ and $Y$ described by
\begin{equation}\label{varp}
\begin{cases}
x_t&= \phi_0^{1}+\sum_{j=1}^p\phi_j^{11}x_{t-j}+\sum_{i=1}^p\phi_i^{12}y_{t-i}+\epsilon^1_t\\
y_t&= \phi_2^{1}+\sum_{j=1}^p\phi_j^{21}x_{t-j}+\sum_{i=1}^p\phi_i^{22}y_{t-i}+\epsilon^2_t
\end{cases}
\end{equation}
where $\epsilon_t^1,\epsilon_t^2$ are taken to be two uncorrelated white-noise series. The goal of the test \citep{granger1969investigating} is to assess the statistical significance of $\{\phi_i^{12}\}_{i=1,...,p}$ by considering as null hypothesis that they are zero, \ie $H_0\: :\: \{\phi_i^{12}=0\}_{i=1,...,p}$. The null hypothesis $H_0$ is equivalent to considering that $\{x_t\}$ evolves according to a AR(p) process. After estimating both VAR(p) and AR(p) models, an F-test \citep{johnston1972econometric} is applied in order to test if VAR(p) outperforms statistically AR(p) in fitting the observations $\{x_t\}$. If it does, $H_0$ is rejected, meaning that $Y$ `Granger-causes' $X$.  

\paragraph{Granger causality network} Having established how to detect a causal relation, we can consider the network of airports where a link $i\rightarrow j$ is present if $i$ `Granger causes' $j$. This approach has already been considered in some recent works in Econometrics \citep{billio2012econometric,corsi2018measuring} and in a recent analysis of the Chinese air transportation network \citep{zanin2017network}. Given $N$ time series, representing the state of delay of the $N$ airports in the network, Granger causality test is performed on all the possible $M=N(N-1)$ pairs. When performing multiple hypothesis testing, a correction to the significance level of each single test should be applied to obtain the desired overall level $\gamma$, \ie if we test $M$ hypotheses simultaneously with a desired $\gamma$, then a significance level $\gamma'<\gamma$ should be applied to each single test to correct for the increased chance of rare events, and therefore, the increased probability of false rejections \citep{tumminello2011statistically}. This has typically not been considered in the literature. However, it can have a huge impact on the number of detected causal links, as we show in the following. Here, we apply the Bonferroni correction which compensates for this effect in the most conservative way by setting $\gamma'=\gamma/M$. Standard topological network metrics can then be extracted from the network of causal relations, \eg link density, clustering, assortativity, efficiency, diameter, centrality rankings of nodes. Each of these metrics describes some specific structural characteristic of the Granger causality network. For example, link density is a measure of the coupling of airports, since a larger number of links means more delay propagation, while measures of node centrality indicate which airports are participating more often to delay propagation.

\subsubsection{Application of Granger causality metrics to the US flights dataset}
\label{sec:data_caus}
Time series of the state of delay for each airport are built for the period from January $1$st 2015 to March $31$st 2015. As suggested in \citep{zanin2017network}, a Z-Score standardisation procedure is applied to reduce the non-stationarity of the time series caused by daily seasonality, which may result in a biased evaluation of the Granger causality metric. The standardised time series of airport $i$ is calculated as $\tilde{x}_{i,t}=(x_{i,t}-\bar{x}_i^t)/\sigma_i^t$
where $\bar{x}_i^t$ and $\sigma_i^t$ are the mean and the standard deviation of the delay states of airport $i$ recorded at hour $t$ across all available days. Hence, pairwise Granger causality tests are applied to the new standardised time series according to Eq. \ref{varp} for different $p$, ranging from $1$ to $6$ hours. The maximum lag is chosen equal to $6$ because the empirical partial autocorrelation function becomes statistically zero after the sixth lag for the time series of any airport. 
In case of rejection of $H_0$, the best $p$ is selected according to the Bayesian Information Criterion. Best $p$ values are distributed around 1 and 2 hours, meaning that delay propagation happens on short timescales. Finally, we set $\gamma=5\%$ and, as a consequence, the significance level of each test is $\gamma'=\frac{0.05}{N(N-1)}$ where $N=315$. 

\begin{figure}[t]
\begin{center}
\includegraphics[width=90mm]{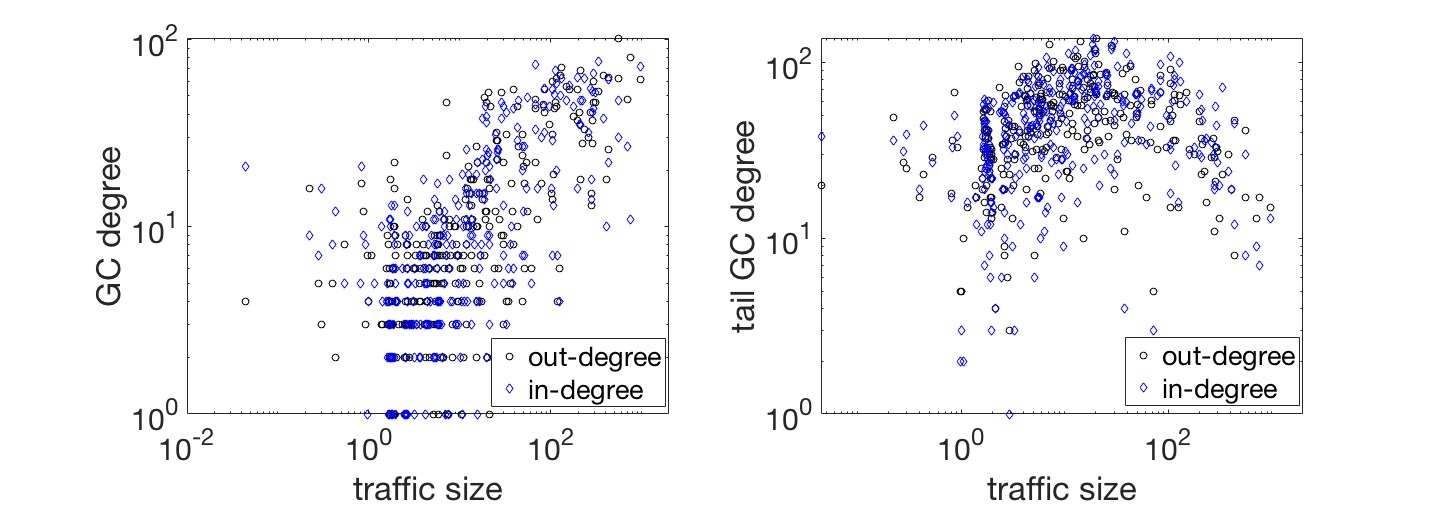}
\includegraphics[width=90mm]{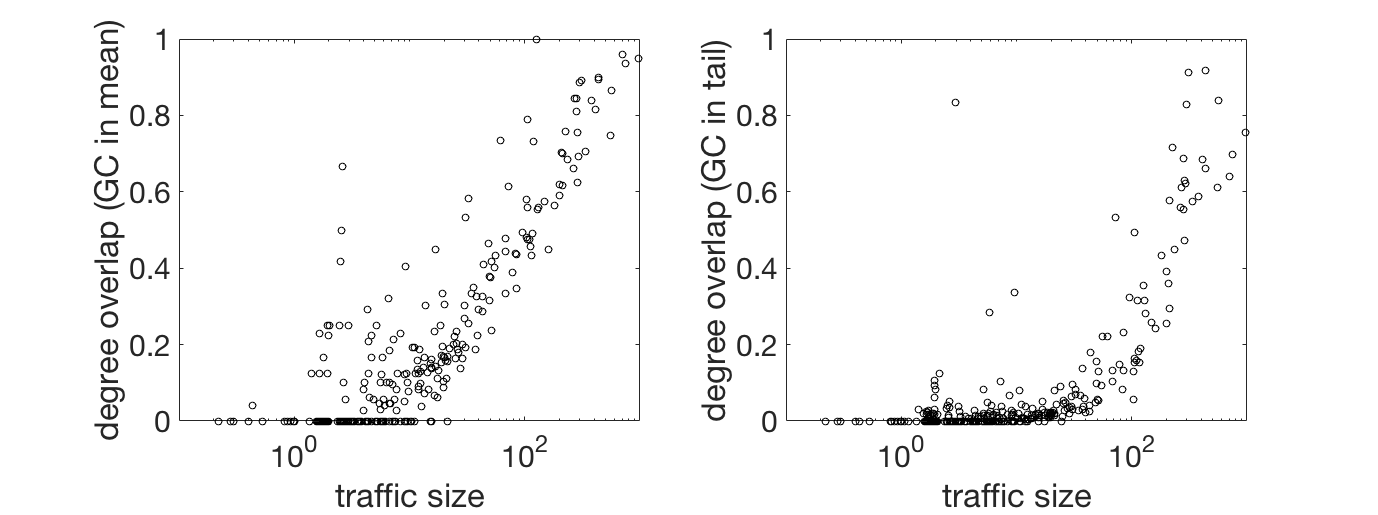}
\caption{Top: degree in the Granger causality networks, both in mean (left) and in tail (right), as a function of the airport traffic size defined as the average number of flights per day. Bottom: degree overlap between the Granger causality networks, for both in mean (left) and in tail (right), and the network of airports and flights described by the adjacency matrix having entry equal to one if there exist flights connecting two airports, zero otherwise. Each point represents the degree overlap averaged over the out-degree and the in-degree of the node.}
\label{fig:GCdegreeAndOverlap}
\end{center}
\end{figure}

The obtained Granger causality network has $L=4401$ Granger causal links. Note that the link density for the Bonferroni corrected network is $\sim 0.04$, whereas without the correction we obtain $\sim0.45$, much larger. Therefore, neglecting to introduce a multiple hypothesis correction means considering a large number of non-significant causal links.
We find a positive linear correlation ($0.62$) between airport size, measured as the average number of flights per day, and node (in- or out-) degree in the Granger network, see the top left panel of Figure \ref{fig:GCdegreeAndOverlap}. The figure shows how the node degree increases (on average) monotonically when the traffic size of the airport increases. Thus, Granger causality in mean suggests that airports having many flights tend to `Granger cause' more than medium-sized and small airports, thus resulting more important in propagating (mean) delays. Furthermore, the channels of propagation are mainly represented by one-leg effects, \ie flights arriving to (for the incoming causal links) or departing from (for the outgoing causal links) the airport. To see this, let us define the degree overlap between the Granger causality network and the network of airports and flights, as
\begin{equation}\label{eq:overlap}
o_i^{out}\equiv \frac{\sum_{j}G_{ij}A_{ij}}{\sum_{j}G_{ij}},~~~~\:o_j^{in}\equiv \frac{\sum_{i}G_{ij}A_{ij}}{\sum_{i}G_{ij}}
\end{equation}
for both the outgoing and the incoming degrees of node $i$ and $j$, respectively, where $G$ ad $A$ are, respectively, the adjacency matrices of the two networks\footnote{ $G_{ij}=1$ if $i$ `Granger causes' $j$, zero otherwise; $A_{ij}=1$ if there exists at least one flight departing from $i$ and arriving to $j$, zero otherwise.}. The degree overlap in Eq. \ref{eq:overlap} measures  how often two airports that are linked in the causality network are also linked in the network of flights. A large degree overlap, therefore, means that a causality link between two airports is often present when the two airports are linked by direct flights. On the contrary, a small degree overlap means that causal relationships are often present even in the absence of a direct flight. Hence, the degree overlap can be interpreted as an indirect measure of the fraction of one-leg effects as channels of delay propagation. It is interesting to notice, see the bottom left panel of Figure \ref{fig:GCdegreeAndOverlap}, that the degree overlap increases with the airport traffic size (Kendall rank correlation equal to $0.65$), and it is very close to one for the largest airports. This is a signal that the primary channels of (mean) delay propagation for large airports are the one-leg effects.

We find also that the diameter of the Granger network, \ie the longest path connecting two nodes, is equal to $8$ while for an Erdos-Renyi random graph with the same number of links (on average) is $4$, thus suggesting the presence of outlying nodes less connected with the central core. This is confirmed also by the average path length, equal to $3.05$ in the Granger network and to $2.4$ in the corresponding Erdos-Renyi network.
The clustering coefficient of a graph is a measure of the likelihood that nodes cluster together, specifically it is the number of closed triangles, \ie subgraphs of three nodes connected each other by links having any direction, divided by the number of any open and closed triangle. The clustering coefficient is $0.28$ in the Granger network, a number much larger than the one of the corresponding Erdos-Renyi network ($0.08\pm 0.01$). This  difference is explained by the different degree of nodes. In fact, the fitness model \citep{caldarelli2002scale}, which preserves on average the degree sequence, has a global clustering coefficient of $0.29\pm0.01$, in line with the empirical Granger network. However, when we consider only feedback triangles, \ie triangles with all links directed clockwise (or anti-clockwise), among all possible triplets, we count $14,856$ such triangles, a number much larger than the corresponding random cases, $908\pm 46$ for the Erdos-Renyi network and $7,656\pm 352$ for the fitness model, suggesting that these feedback loops are over-expressed in the ATM system. In fact, a feedback triangle represents a positive feedback subsystem which tends to amplify delay propagation, thus making the system more unstable. Another subsystem for delay amplification is represented by a reciprocated link between two nodes. Reciprocity is a measure of the likelihood that nodes in a directed network are mutually linked and the reciprocity coefficient is defined as the ratio of the number of links pointing in both directions to the total number of links. In the empirical Granger network the reciprocity is $0.20$, a value larger than $0.02\pm0.01$ for the Erdos-Renyi network and $0.09\pm 0.01$ for the fitness model.

Hence, in the case of ATM systems, interesting network metrics are the ones which considers feedback loops or reciprocal links and, any innovation which aims to increase the resilience of the system to delay propagation should tend to reduce them.

Moving to node-specific topological metrics allows us to better characterise the US ATM system. In particular, PageRank centrality applied to the causal network reveals that the most important airports in the propagation of (mean) delays are the ones with high traffic, such as Orlando, Atlanta, and Charlotte, to name but the top three. That is, large airports are more informative regarding the prediction of the state of delay of the whole system and more central for the process of delay propagation. This finding, however, contradicts the conclusions of \citep{zanin2017network} which, on the contrary, points out the centrality of small and regional airports for the propagation of delays in the Chinese air transport system.

Finally, by repeating the pairwise causality analysis for a time window of one month and rolling the window week-by-week, we notice that link density, \ie a measure of how much the system is interconnected, changes significantly also when aggregated quantities, such as the total traffic or the mean delay, are quite constant in the considered period. This is a signal of a complex dynamics of delay propagation, which is not simply explained by the total traffic in the air system.

\subsubsection{Granger causality in tail} 
\label{sec:new_metrics_caus}
The results presented in the previous section are based on linear models. However, the complex nature of the delay propagation dynamics might not be fully captured by a linear analysis. For example, departing delays which are small with respect to flight time are probably not relevant for delay propagation, as they are easily absorbed during the flight or by buffers. Small states of delay of airports are nevertheless considered by the Granger causality in mean test, which weights equally small and large values in assessing the statistical significance of the past information of $Y$ in forecasting $X$, see Eq. \ref{varp}. 
For this reason, we propose to use an extension of the Granger causality test, namely \emph{Granger causality in tail} \citep{hong2009granger}, which considers only extreme events, defined as states of delay falling in the right tail of the distribution, \ie large delays. With the same spirit of \citep{granger1969investigating}, Granger causality in tail aims to evaluate whether extreme events in one airport cause extreme events in another airport. An extreme event for the state of delay of an airport is thus interpreted as a state of congestion for that airport. The Granger causality in tail test works as follows. Assume to know at each step the probability density function of $X$ conditional on past values\footnote{Conditional density for a time series can be estimated, \eg by  historical simulation methods or autoregressive conditional density model \citep{hansen1994autoregressive}. } and let us define $V_t\equiv V(x_1,...,x_{t-1},\beta)$ as the $(1-\beta)$-quantile of the conditional probability distribution of the time series $X$, \ie $\mathbb{P}(X>V_t|x_1,...,x_{t-1})= 1-\beta$ almost surely with $\beta\in(0,1)$ defines $V_t$ implicitly.
The null hypothesis $H_0^{tail}$ of \citep{hong2009granger} is:
\begin{equation}\label{tailGranger}
\mathbb{P}(X>V_t|\{x_s\}_{s=1}^{t-1})=\mathbb{P}(X>V_t|\{x_s\}_{s=1}^{t-1},\{y_s\}_{s=1}^{t-1}) \:\mbox{ a.s.}
\end{equation}
meaning that predicting an extreme event for $X$ with or without the past information on $Y$ is statistically equivalent. the alternative hypothesis is that the above probabilities are different. Thus a rejection of the null hypothesis $H_0^{tail}$ means that $Y$ `Granger causes in tail' $X$ at level $\beta$. Ref. \citep{hong2009granger} derives under the null hypothesis a Central Limit Theorem for a suitably standardized sum over lags of the squares of the sample cross-correlations of the binary variables where value $1$ is assigned to a tail events. This theorem allows to extract a p-value associated with the null hypothesis. For further information on how to make testable the definition in Eq. \ref{tailGranger}, see \citep{hong2009granger}. In the analysis of the US data, we adopt the autoregressive conditional density model \citep{hansen1994autoregressive} to characterise the conditional probability distribution, by assuming an AR(p) model for $X$ with i.i.d. Gaussian innovations and $\beta=0.05$.

We apply the pairwise Granger causality in tail test to the standardised time series of the  state of delay. The dataset and the test p-value and correction are the same used in the study of Granger in mean.
The obtained Granger causality in tail network has $L=15,027$ causal links, thus link density for the Bonferroni-corrected network is $\sim 0.15$, quite larger than the `in mean' one, suggesting that restricting to the extreme delays is much more informative that considering delays of all sizes. Comparing the two causality networks, we find that around half ($\sim 0.46$) of the causal links present in the `in mean' network are also present in the `in tail' network. The differences between the 'in mean' and 'in tail' networks are due to the presence, in the latter, of less causal links associated with large airports and more causal links associated to small and medium airports, see the top right panel of Figure \ref{fig:GCdegreeAndOverlap}. This difference is further confirmed by the low Spearman correlation between the node degree of the Granger causality in mean network and the corresponding one in the `in tail' case, \ie $0.20$ when considering the out-degree and $0.32$ for the in-degree. The non-monotonic behaviour of both the out- and in-degree in the Granger causality in tail network as a function of the traffic size of airports observed in Figure \ref{fig:GCdegreeAndOverlap} is a signal of the importance of small and medium-size airports in the process of propagation of extreme delays.
It is interesting to notice that we measure a positive rank correlation (Kendall coefficient $0.56$) between the degree overlap (computed according to Eq. \ref{eq:overlap}) for the Granger causality in tail network and the airport size (in terms of traffic). Again, the overlap is close to one for the largest airports, a signal of one-leg effects as propagation channels for those airports. On the other hand, it is close to zero for both small and medium-size airports, suggesting that the mechanisms of delay propagation are represented, in this case, by two or more legs effects. In other words, a channel of delay propagation from a small airport to another airport occurs by means of two or more flights which create a path connecting them by involving other airports in between\footnote{In principle, other exogenous sources may be responsible for the presence of a causal relationship, \eg
weather may create a correlation between the states of delay of two airports that are geographically close.
Thus, a dependence between two states of delay might also not be due to flights.}.

Similarly to the Granger causality in mean case, we find that some standard network metrics are over-expressed with respect to the corresponding random cases. In particular, the average path length is equal to $1.95$ for the Granger in tail network, slightly larger than $1.84\pm 0.01$ corresponding to the Erdos-Renyi case, but close to the value $1.90\pm 0.01$ for the fitness model, thus highlighting that the average path length can be explained in terms of degree distributions of the nodes. A similar behaviour is observed for the clustering coefficient, equal to $0.26$ for the causality network, $0.16\pm 0.01$ for the Erdos-Renyi case, and $0.25\pm 0.01$ for the fitness model. However, both feedback triplets ($71,127$) and the reciprocity coefficient ($0.14$) are significantly over-expressed with respect to the random cases represented by both Erdos-Renyi and fitness models, respectively showing $3.631\pm 871$ and $64.136\pm 1.415$ feedback triplets and a reciprocity coefficient of $0.07\pm 0.01$ and $0.11\pm 0.01$. This result confirms further that the over-expression of feedback loops and reciprocal links in the causality networks is a characteristic property of ATM systems. 

Finally, in the Granger causality in tail network the most central airports according to PageRank are different from the ones selected by Granger causality in mean and, more specifically, are characterised by low traffic. Hence, this result indicates that extreme delays are mostly propagated from small and regional airports. 

\subsection{Comparing centrality and causality}
When ranking airports by Trip centrality loss, the highest rank airports are the ones with smaller losses, i.e. the ones for which outgoing (or incoming) itineraries were more preserved. Therefore, we expect that the highest ranked airports are those such that their outgoing (or incoming) flights are less delayed, therefore causing less itineraries disruption due to missed connections. Given that the degree of an airport (outgoing or incoming) in the Granger causality network measures to how many airports it propagates delay (or how many airports propagate delay to it), we expect that airports with large causality degree tend to have a large centrality loss, i.e. that the two rankings have a negative correlation.  
\begin{figure}[t]
\begin{center}
\includegraphics{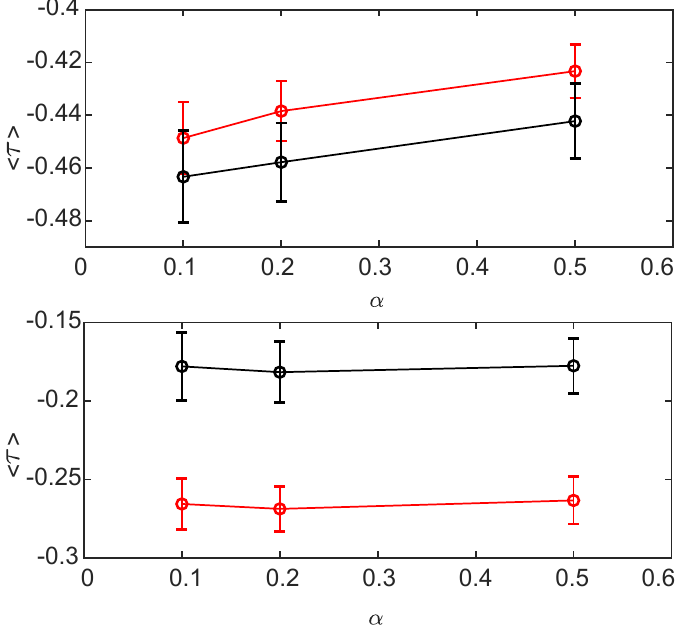}
\caption{Kendall correlation coefficient between the ranking according to Trip Centrality loss and that according to the degree in the Granger causality network (upper panel: in mean; lower panel: in tail) for different values of $\alpha$ used to compute Trip centrality. The coefficient is computed between the ranking obtained with Trip Centrality loss on each day of April (the airport losing lass centrality is ranked first) and the unique ranking obtained from the degree on the causality network, and then averaged over all days.  Bars represent standard deviations. The red line corresponds to the incoming centralities and the black one to the outgoing.}
\label{fig:tc_gc}
\end{center}
\end{figure}
This is actually what we observe in Figure \ref{fig:tc_gc}, where we plot the relation between rankings according to centrality loss and Granger causality as a function of the parameter $\alpha$ of Trip centrality. 
For small $\alpha$s, \ie when more weight is given to short trips in the computation of Trip centrality, the ranking of airports according to Trip centrality loss and Granger causality in mean are strongly inversely correlated (see the top panel of Figure \ref{fig:tc_gc})  This is in part explained by the fact that both the loss of centrality and the degree in the causality network tend to be larger for airports with higher traffic\footnote{See the top left panel of Figure \ref{fig:GCdegreeAndOverlap} to notice this behaviour for Granger causality in mean.}. However, when $\alpha$ is increased, \ie  when longer trips are weighted more in the computation of Trip centrality, the ranking according to centrality loss changes in a complex way depending on the itineraries of more than one leg (see \cite{Zaoli2019}), thus reducing the negative correlation with the ranking according to Granger causality in mean, which tends to capture one-leg effects, especially for large airports. Finally, we notice in the bottom panel of Figure \ref{fig:tc_gc} a negative correlation between Trip centrality loss and Granger causality in tail, but weaker than for the Granger in mean. This is somehow expected since Granger causality in tail highlights the importance (high degree in the causality network) of peripheral airports in the propagation of extreme delays, but the same airports are less important in terms of trips, thus they have little Trip centrality and consequently also little Trip centrality loss.

\section{Conclusions}
\label{sec:conclusions}
In this paper we presented the toolbox 
proposed by the Domino project to assess the system-wide impacts of of introducing new mechanism into the ATM system. 

The proposed toolbox consists of network metrics capable of detecting the effects of the changes on the interaction of the network elements. In particular, centrality and causality metrics have been considered, owing to their capacity to measure the network connectivity and the propagation of delays and congestion in the network.  However, we have shown here that existing centrality and causality metrics are not sufficient to describe the ATM system. Specifically, existing centrality metrics are not able to tell apart a situation where delays disrupt important connections to one where they do not and do not account in a satisfactory way for the multiplex nature of the network. On the other hand, commonly used causality metrics assume linearity in the delay propagation, which might not be realistic. We therefore introduced new centrality and causality metrics, whose functioning we illustrated on a dataset of US flights. 

In a separate forthcoming paper we will show how to use the proposed metrics to assess the effect of introducing specific innovations in ATM. This is possible thanks to the Domino Agent Based Model which, once carefully calibrated on real air traffic data, allows to simulate the whole European airspace in the current scenario as well as in future scenarions modified by the introduction of new mechanisms.\\

\section*{Acknowledgements}
This project has received funding from the SESAR Joint Undertaking under grant agreement No 783206 under European Union’s Horizon 2020 research and innovation programme. The opinions expressed herein reflect the authors’ views only. Under no circumstances shall the SESAR Joint Undertaking be responsible for any use that may be made of the information contained herein.


\end{document}